\documentclass[aps,prl,twocolumn,showpacs,groupedaddress,bibnotes]{revtex4}

\usepackage{graphicx}% Include figure files
\usepackage{bm}% bold math

\newcommand\pictc[5]{\begin{figure}
            \centerline{\vspace{0mm}
\includegraphics[width=#1\columnwidth,height=0.7\textheight,keepaspectratio]{#3}}
            \protect\caption{\protect\label{fig:#4} #5}%\vspace{-2mm}
                    \end{figure}            }

\newcommand\pict[4][1.0]{\pictc{#1}{!tb}{#2}{#3}{#4}}
\newcommand\rpict[1]{\ref{fig:#1}}

\newcommand\leqt[1]{\protect\label{eq:#1}}
\newcommand\reqtn[1]{\ref{eq:#1}}
\newcommand\reqt[1]{(\reqtn{#1})}

\newcounter{Fig}

\begin{document}
\begin{sloppy}

\title{Defect-free surface states in modulated photonic lattices}

\author{Ivan L. Garanovich}
\author{Andrey A. Sukhorukov}
\author{Yuri S. Kivshar}

\affiliation{Nonlinear Physics Centre and Centre for Ultra-high bandwidth
Devices for Optical Systems (CUDOS),
 Research School of Physical Sciences and Engineering,
 Australian National University, Canberra, ACT 0200, Australia}

\begin{abstract}
We predict
that interfaces of modulated photonic lattices can support a novel type of
generic surface states. Such linear surface states appear in  truncated but otherwise perfect (defect-free) lattices as a direct consequence of the periodic modulation of the lattice potential, {\em without} any embedded or nonlinearity-induced defects. This is in a sharp contrast to all previous studies, where surface states in linear or nonlinear lattices, such as Tamm or Shockley type surface states, are always associated with the presence of a certain type of structural or induced surface defect.
\end{abstract}



\maketitle
Interfaces separating different physical media can support a special
class of transversally localized waves known as {\em surface waves}.
Linear surface waves have been studied extensively in many branches
of physics~\cite{Davidson:1996:SurfaceStates}.
For example, electro-magnetic waves localized at the boundaries of periodic photonic structures, such as waveguide arrays or photonic crystals, have been extensively analyzed theoretically and experimentally.
The appearance of localized surface waves in photonic structures is commonly explained as the manifestation of {\em Tamm or Shockley type localization mechanisms}~\cite{Tamm:1932-849:ZPhys, Shockley:1939-317:PREV, Malkova:2007-45305:PRB}, being {\em associated with the presence of a certain type of surface defect}.
Tamm states were first identified as localized electronic states at the edge of a truncated periodic
potential~\cite{Tamm:1932-849:ZPhys}, and then they were found in other systems, e.g. at an interface separating periodic and homogeneous dielectric optical
media~\cite{Yeh:1977-423:JOS, Yeh:1978-104:APL}.

In discrete systems, such as arrays of weakly coupled optical waveguides~\cite{Christodoulides:2003-817:NAT},
different types of linear and nonlinear states localized at and near the
surface have also been analyzed extensively.
It was found that Tamm surface waves can exists at the edge of an array of optical waveguides when the effective refractive index of the boundary waveguide is modified above a certain threshold~\cite{Makris:2005-2466:OL, Molina:2005-35404:PRB, Suntsov:2006-63901:PRL, Kartashov:2006-73901:PRL, Siviloglou:2006-5508:OE, Smirnov:2006-2338:OL, Rosberg:2006-83901:PRL}, whereas surface localization was considered to be impossible when all waveguides are exactly identical, as sketched in Fig.~\rpict{straight_vs_curved}(a). In the latter case, the beam launched into array delocalizes due to diffraction [Fig.~\rpict{straight_vs_curved}(b)], and it is also strongly reflected from the boundary as illustrated in Fig.~\rpict{straight_vs_curved}(c).

In this Letter {\em we predict,} for the first time to our knowledge and contrary to the accepted notion, that {\em novel type of generic defect-free surface waves can exist at the boundary of a periodic array of identical optical waveguides,} which axes are periodically curved along the propagation direction as schematically shown in Fig.~\rpict{straight_vs_curved}(d).
The periodic bending of waveguide axes was shown to result in the modification of diffraction~\cite{Eisenberg:2000-1863:PRL, Longhi:2006-243901:PRL, Garanovich:2006-66609:PRE, Iyer:2007-3212:OE}, which strength nontrivially depends on the waveguies bending and optical wavelength. An interesting feature is that the diffraction can be completely suppressed for particular values of the bending amplitude, and this effect is known as dynamic localization or beam self-collimation~\cite{Eisenberg:2000-1863:PRL, Longhi:2006-243901:PRL, Garanovich:2006-66609:PRE, Iyer:2007-3212:OE}. Under such very special conditions, the beam experiences periodic self-imaging, propagating without spreading for hundreds of free-space diffraction lengths, as illustrated in Fig.~\rpict{straight_vs_curved}(e). On the other hand, if the beam is launched at the edge of a {\em semi-infinite} modulated lattice tuned to the self-collimation, one can intuitively expect that it can not penetrate deep into the lattice as away from the lattice edge effect of the boundary is negligible and coupling between the lattice sites is canceled in the self-collimating lattice. In Fig.~\rpict{straight_vs_curved}(f) we indeed observe
that the beam remains localized at the surface of the self-collimating modulated lattice. However, our most nontrivial finding detailed below is that {\em surface localization is possible for an extended range of structural parameters even when diffraction is non-vanishing}.

\pict{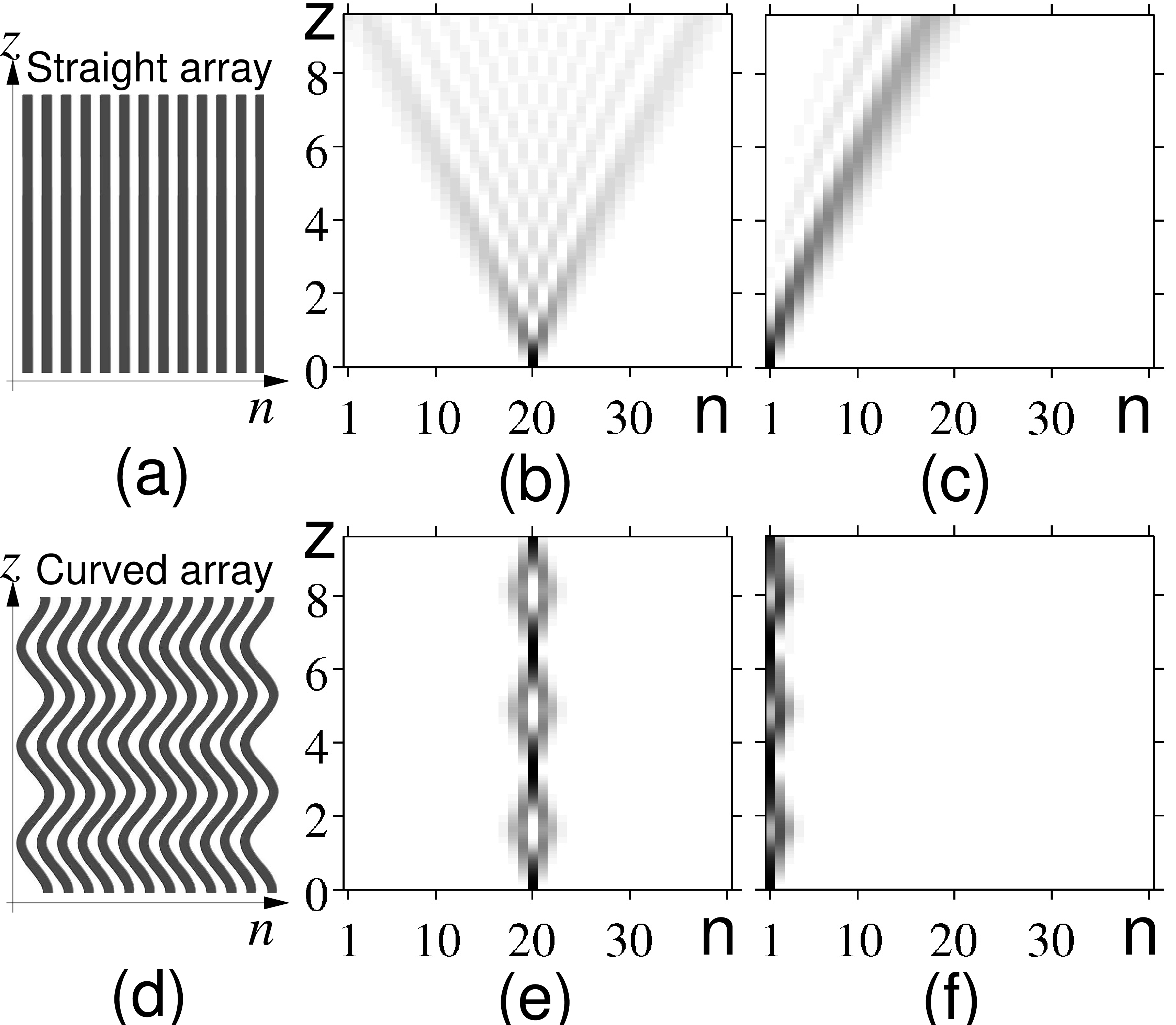}{straight_vs_curved}{
(a-c)~Beam propagation in a straight semi-infinite lattice shown schematically in (a): (b)~discrete diffraction away from the lattice boundary and (c)~diffraction and reflection from the surface when beam is coupled to a boundary waveguide.
(d-e)~Beam dynamics in a sinusoidally modulated lattice, shown schematically in (d). The modulation amplitude is chosen to obtain self-collimation regime (${\rm A = A}_0 \simeq 1.24$ for the period ${\rm L} = 3.25$). Diffraction is suppressed and beam remains localized at the input location either (e)~away or (f)~at the lattice boundary.
}

We study propagation  and localization of light in a semi-infinite one-dimensional array of coupled optical waveguides, where the waveguide axes are periodically curved in the propagation direction $z$ with the period  ${\rm L}$, as shown schematically in Fig.~\rpict{straight_vs_curved}(d). When the tilt of beams and waveguides at the input facet is less than the Bragg angle, the beam propagation is primarily characterized by coupling between the fundamental modes of the individual waveguides, and it can be described by the tight-binding equations taking into account the periodic waveguide bending~\cite{Longhi:2006-243901:PRL, Longhi:2005-2137:OL},
\begin{equation} \leqt{DNLS}
   {\rm i} \frac{{\rm d a_{n}}}{{\rm d} z}
   + {\rm C } \exp\left[ {\rm -i \dot{x}}_0(z) \right] {\rm a_{n+1}}
   + {\rm C } \exp\left[ {\rm  i \dot{x}}_0(z) \right] {\rm a_{n-1}}
   = 0 ,
\end{equation}
where ${\rm a_n(}z{\rm )}$ is the field amplitude in the ${\rm n}$-th waveguide,
${\rm n} = 1, \ldots$, and ${\rm a_{n \le 0}} \equiv 0$ due to the structure
termination. Transverse shift ${\rm x}_0(z) \equiv {\rm x_0(}z+{\rm L})$ defines the periodic longitudinal lattice modulation. Coefficient ${\rm C}$ defines the coupling strength between the neighboring waveguides, it characterizes diffraction in a straight waveguide array with ${\rm x}_0 \equiv 0$ \cite{Somekh:1973-46:APL} [see an example  in Fig.~\rpict{straight_vs_curved}(b)]. Expression~\reqt{DNLS} shows that the effect of periodic lattice modulation appears through the modifications of phases of the coupling coefficients along the propagation direction $z$. In order to specially distinguish the effects due to diffraction management, we consider the light propagation in the waveguide arrays with symmetric bending profiles, since asymmetry may introduce other effects due to the modification of refraction, such as beam dragging and steering~\cite{Kartashov:2005-1378:OL, Garanovich:2005-5704:OE, Rosberg:2006-1498:OL}. Specifically, we require that ${\rm x}_0(z) = {\rm f}(z - \tilde{z})$ for a given coordinate
shift $\tilde{z}$, where function ${\rm f}(z)$ is symmetric, ${\rm f}(z) \equiv {\rm f}(- z)$.

In order to analyze light propagation near the surface of a semi-infinite modulated lattice, we first consider the case of small modulation periods ${\rm L}$, such that the parameter $\kappa = 2\pi/{\rm L}$ is large, $\kappa \gg 1$. Then we can employ the
asymptotic expansion (see, e.g., Ref.~\cite{Kivshar:1994-2536:PRE})
${\rm a_n}(z) = {\rm u_n}(z)
              + \sum_{{\rm m}\ne 0} {\rm v_{n,m}}(z)\exp({\rm i m} \kappa z)$,
where ${\rm u_n}(z)$ have the meaning of the averaged field values over the modulation period, and we take into account first- and second-order terms for the oscillatory corrections which have zero average, ${\rm v_{n,m}} = {\rm v_{n,m}}^{(1)} \kappa^{-1} + {\rm v_{n,m}}^{(2)} \kappa^{-2}$.
Since the modulation is periodic, we can perform Fourier expansion of the coupling coefficients as ${\rm C} \exp[{{\rm -i \dot{x}}_0(z)}] = \sum_{\rm m} C_{\rm m} \exp({\rm im} \kappa z)$. Then, in the regime close to self-collimation when the average coupling is small, $|{\rm C}_0| \sim O(\kappa^{-1})$, we combine the terms of the same orders~\cite{Kivshar:1994-2536:PRE} and finally
obtain the effective equations for the slowly varying functions ${\rm u_n}(z)$,
\begin{equation}
   \begin{array}{l} {\displaystyle
        {\rm i} \frac{{\rm d u_{n}}}{{\rm d} z}
        + {\rm C}_0 {\rm u_{n+1}}
        + {\rm \bar{C}}_0 {\rm u_{n-1}}
   } \\*[9pt] \qquad {\displaystyle
        + \delta_{1,{\rm n}}\left(\Delta_1{\rm u_1}+\Delta_2 {\rm u_2}\right)
        + \delta_{2,{\rm n}}\bar{\Delta}_2{\rm u_1}
        = 0.
   } \end{array}
\end{equation}
Here $\delta$ is the Kronecker delta,
$$\Delta_1 = - \kappa^{-1} \sum_{{\rm m} \ne 0} |{\rm C}_{\rm m}|^{2} {\rm m}^{-1},$$
$$\Delta_2 = \kappa^{-2} \sum_{{\rm m} \ne 0}\sum_{{\rm j} \ne 0,{\rm -m}}
 {\rm C}_{\rm j} {\rm C}_{\rm m} \bar{\rm C}_{\rm j+m} {\rm j}^{-1} {\rm m}^{-1},$$
the bar stands for the complex conjugation, and ${\rm u_{n \le 0}} \equiv 0$.
From these equations one can see that the effect of periodic modulation is to introduce the
''virtual'' defects $\Delta_1$ and $\Delta_2$ at the lattice boundary.
We now seek solutions in the form of stationary modes, ${\rm u_n}(z) = {\rm u_n}(0) \exp( {\rm i k} z / {\rm L })$, where ${\rm k}$ is the Bloch wave-number.
The values of $|\rm k| \le 2 |{\rm C}_0| L$ correspond to a {\em transmission band}, where the modes are infinitely extended. On the other hand, the modes can become localized at the surface of semi-infinite modulated lattices is $|\rm k| > 2 |{\rm C}_0| L$, and we find that such solutions exist if the modulation parameters are sufficiently close to the self-collimation condition where $|{\rm C}_0|$ is small. Specifically, there exists one surface state if $\alpha_2 - \alpha_1 \leq 2 \leq \alpha_1 + \alpha_2$, and two surface states emerge if
$\alpha_2 - \alpha_1 > 2$, where $\alpha_1 = |\Delta_1/{\rm C}_0|$ and $\alpha_2 = |1 + \Delta_2/{\rm C}_0|^{2}$.

We note that {\em if the modulation is symmetric}, such that $\dot{{\rm x}}_0(z + {\rm L}/2) = -\dot{{\rm x}}_0(z)$, then $|{\rm C_m}| \equiv |{\rm C}_{\rm -m}|$ and accordingly $\Delta_1 = 0$, meaning that the modes should always appear in pairs. Moreover, this conclusion is valid even beyond the applicability of the asymptotic expansion, since we identify the exact symmetry of the model Eq.~\reqt{DNLS} in case of symmetric modulations: for each solution ${\rm a_n}(z)$,
${\rm b_n}(z) = (-1)^{\rm n} \bar{{\rm a}}_{\rm n}(z + {\rm L}/2)$ is also a solution. Therefore, in symmetric structures surface modes always appear in pairs with the Bloch wavenumbers of the opposite sign, ${\rm k} = \pm {\rm L|C_0|}(d + d^{-1})$, where
$d = \left( |1 + \Delta_2/{\rm C_0}|^{2} - 1 \right)^{1/2}$,

As an example, we further consider a sinusoidal modulation function of the form ${\rm x}_0(z) =  {\rm A} \left[ \cos\left(2\pi z / {\rm L}\right) - 1 \right]$, similar to the one which has recently been employed to demonstrate dynamical localization in modulated waveguide
arrays~\cite{Longhi:2006-243901:PRL, Longhi:2005-2137:OL}. In this case,
the Fourier coefficients can be calculated analytically, ${\rm C_m = C J_m[\xi A/A_0]}$, and ${\rm J}_{\rm m}$ is the Bessel function of the first kind of order ${\rm m}$. The modulation amplitude ${\rm A_0}$ corresponds to the self-collimation condition~\cite{Dunlap:1986-3625:PRB, Longhi:2006-243901:PRL}, $ {\rm A_0} = \xi {\rm L} / (2 \pi)$, where $\xi \simeq 2.405$ is the first root of the Bessel function ${\rm J}_0$. Since the sinusoidal modulation is symmetric, then for each modulation amplitude ${\rm A}$ such that ${\rm A}_{\rm crit}^{-} < {\rm A} < {\rm A}_{\rm crit}^{+}$, where
${\rm A}_{\rm crit}^{-}$ and ${\rm A}_{\rm crit}^{+}$ are the left and right mode cut-offs, respectively, there exists (at least) a pair of surface modes. We use our asymptotic analysis to estimate the cut-off values in the case of small modulation periods,
${\rm A}_{\rm crit}^{\pm}/{\rm A}_0 \simeq 1 \pm 2 {\rm A}_0^{2} {\rm C}_{2} {\rm J}_1[\xi]
{\rm J}_2[\xi]/(\xi^{3}(\sqrt{2}\mp 1))$.

\pict{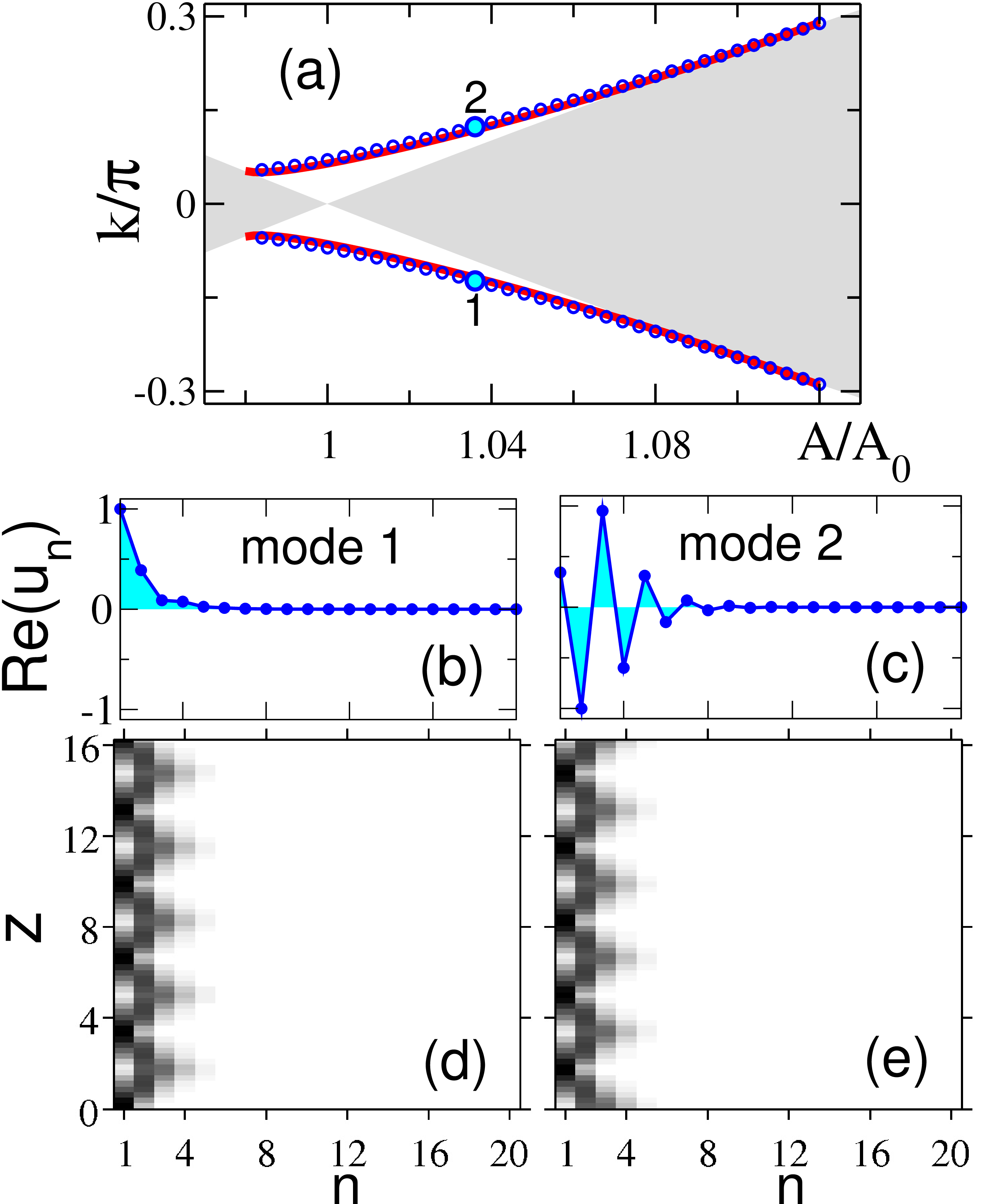}{modes}{
(Color online) (a) Defect-free surface modes in sinusoidally modulated lattice with modulation
period ${\rm L} = 3.25$. Circles and solid lines show modes Bloch wave numbers ${\rm k}$ calculated
numerically  and using asymptotic expansion, respectively. Shading marks transmission band of
the lattice. (b-c)~Numerically calculated modes profiles at the input, and (d-e) their propagation dynamics are
shown for the two complementary modes marked 1 and 2 in (a), respectively.}

In order to confirm our analytical results, we calculate numerically the mode spectrum of the original Eq.~\reqt{DNLS}.
In Fig.~\rpict{modes}(a) one can see that for sufficiently small modulation periods ${\rm L}$
there indeed exists a pair of symmetric surface modes outside the lattice transmission band, and the wave numbers of surface modes calculated using asymptotic expansion are in excellent agreement with those calculated numerically. At the cross-section $z=0$, one mode has unstaggered input profile [Fig.~\rpict{modes}(b)], while the other one exhibits staggered structure [Fig.~\rpict{modes}(c)]. We note that there is very weak additional phase modulation, ${\rm Im[a_n]/Re[a_n]} \sim 10^{-3}$, in agreement with the asymptotic analysis predicting real profiles up to second-order corrections.
In all the figures, we put ${\rm C = 1}$, since results can be mapped to the other coupling values using a simple transformation ${\rm a_n}(z, {\rm C, L, A}) = {\rm a_n(C}z, {\rm C \equiv 1, CL, CA})$.

\pict{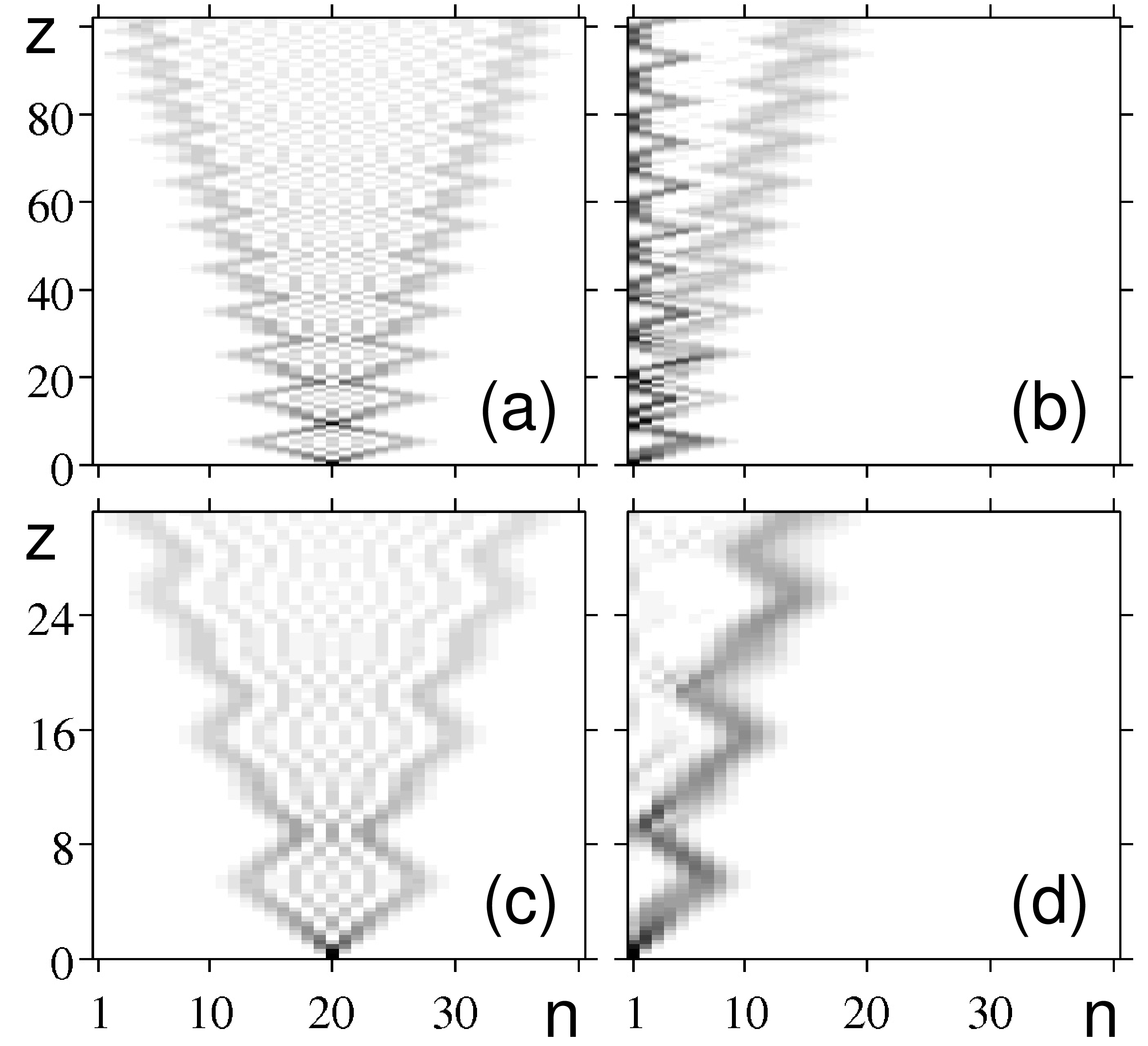}{generation}{
Beam propagation in two modulated semi-infinite lattices with two different modulation amplitudes and the same modulation period ${\rm L} = 9.75$ . Top:~Lattice modulation is larger than the left surface modes cut-off, ${\rm A} \simeq 0.934 {\rm A}_0$. (a) Beam diffracts when launched away from the surface. (b) Surface wave is generated when the edge
waveguide is excited. Bottom:~Modulation amplitude is less than the left surface modes cut-off, ${\rm A} \simeq 0.8 {\rm A}_0$. Beam always diffracts, whether it is launched (c) inside the lattice or (d) at the surface.}

We further demonstrate that defect-free surface modes in modulated lattices can be effectively generated using single-site excitation of the edge lattice waveguide, if the lattice modulation amplitude ${\rm A}$ is between the left and the right cut-offs
${\rm A}_{\rm crit}^{-}$ and ${\rm A}_{\rm crit}^{+}$. An example of such surface wave excitation after some initial radiation is shown in Fig.~\rpict{generation}(b), where even though the lattice modulation is very close to the left cut-off,
${\rm A} \simeq 1.0065 {\rm A}_{\rm crit}^{\rm -}$, the surface wave is still very well localized. In contrast, when the beam is launched far away from the surface, it always diffracts if ${\rm A \ne A_0}$, as shown in Fig.~\rpict{generation}(a). This illustrates the fundamental difference between the dynamical localization in infinite modulated
lattices~\cite{Longhi:2006-243901:PRL, Dunlap:1986-3625:PRB}, and formation of the defect-free surface modes in truncated modulated lattices. While dynamical localization is a purely resonant effect which takes place just for one {\em single} value of the modulation amplitude ${\rm A = A_0}$
[see Fig.~\rpict{straight_vs_curved}(e)], the  {\em families} of defect-free surface modes always exist in a finite range of the modulation amplitudes sufficiently close to the self-collimation value ${\rm A_0}$. If the deviation of the modulation amplitude from the self-collimation value
is greater than the one determined by the left and the right cut-offs, the defect free modes disappear, and the beam always diffracts irrespectively of its input position in the semi-infinite modulated lattice, see  Figs.~\rpict{generation}(c) and~\rpict{generation}(d).

\pict{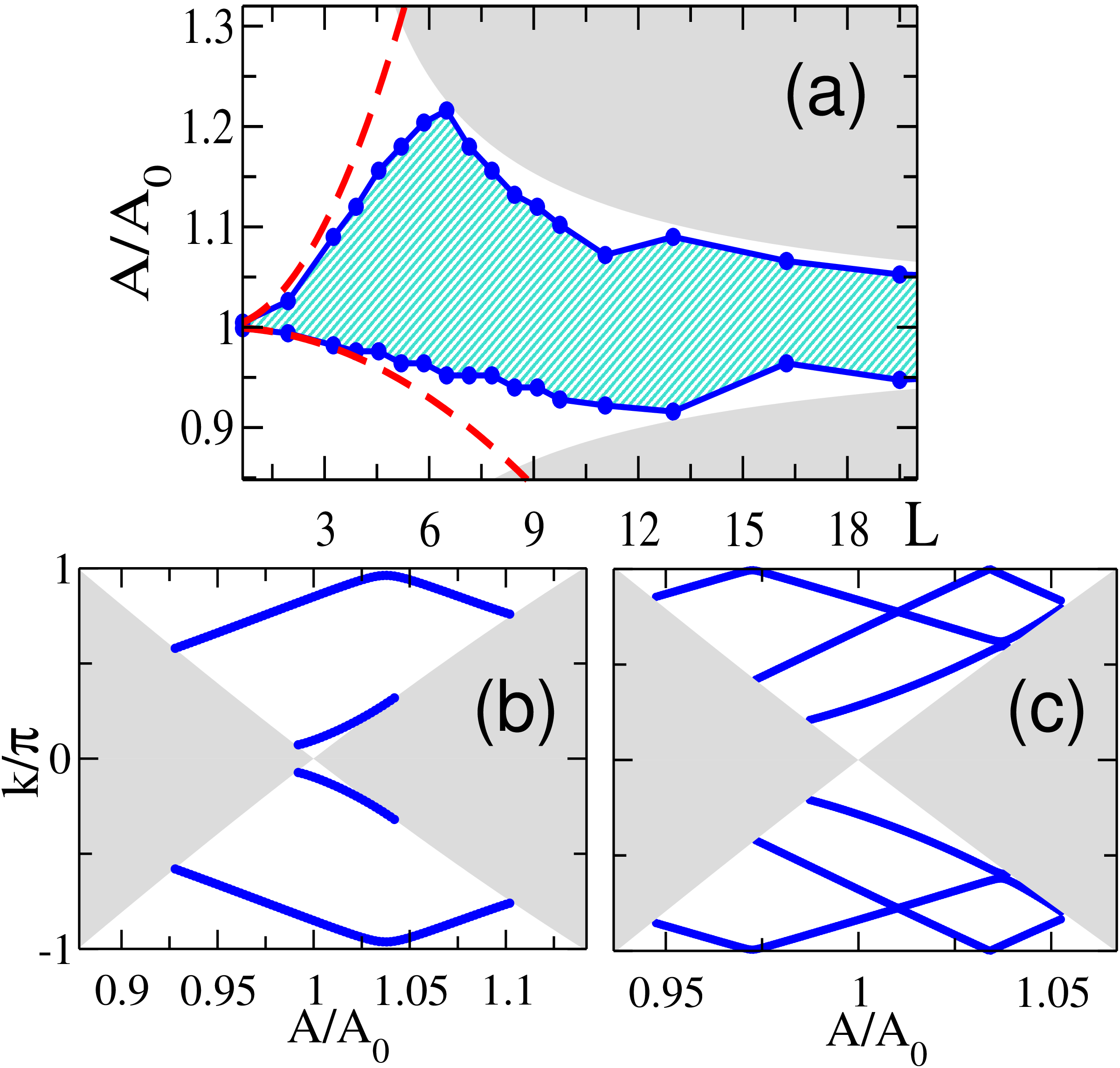}{spectra}{
(Color online) (a) Solid lines bound domain [hatched] of existence of defect-free surface modes
in the sinusoidal modulated lattice calculated numerically. Dashed lines show modes
cut-offs obtained from the asymptotic expansion. Solid shading marks the region where localized modes can not exist. (b-c) Families of defect-free
surface modes for the modulation periods  ${\rm L} = 9.75$ and ${\rm L} = 19.5$, respectively.
Shadings mark lattice transmission bands.}

For large modulations periods the asymptotic analysis is not valid, and we use numerical simulations to find families of the defect-free surface modes. These results are summarized in Fig.~\rpict{spectra}(a), where hatched is the domain of the existence of the defect-free surface modes on the
${\rm (L, A)}$ parameter plane. For small modulation periods, the asymptotic expansion provides an estimate for the surface modes cut-offs (dashed lines).
When the modulation period grows, the number of defect-free modes increases, as shown in Figs.~\rpict{spectra}(b)~and~(c). For the large modulation periods the domain of the existence of the defect-free surface modes is basically limited by the region where lattice transmission band extends to the whole Brillouin zone from $-\pi$ to $-\pi$, and therefore localized states cannot exist. The region where localized modes
can not exist [shown with solid shading in Fig.~\rpict{spectra}(a)] is given by the relation ${\rm L} \ge \pi / (2 |{\rm C}_0|) = \pi / ({\rm C J_m}[\xi {\rm A/A}_0])$.

We note that although the defect-free surface states were introduced here for modulated photonic lattices, such novel type of surface modes may also appear in other fields where wave dynamics is governed by coupled Schr\"odinger type equations~\reqt{DNLS} with $z$ standing for time. In particular, by introducing special periodic shift of lattice potential it may be possible to observe peculiar surface localization in Bose-Einstein condensates. On the other hand, our results indicate the possibility for novel mechanism of surface localization of charged particles in complex time-varying driving electric fields, for which the possibility of the dynamical localization has been suggested earlier~\cite{Dunlap:1986-3625:PRB}.

In conclusion, we have demonstrated, for the first time to our knowledge, that interfaces of modulated photonic lattices can support a novel type of generic {\em defect-free} surface states. Such surface states appear in truncated but otherwise perfect (defect-free) lattices as a direct consequence of the periodic modulation of the lattice potential, {\em without} any embedded or nonlinearity-induced defects. This is in a sharp contrast to all previous studies, where surface states in linear or nonlinear lattices, such as Tamm or Shockley type surface states, are always associated with the presence of a certain type of surface defect. Using both asymptotic expansion technique and numerical simulations, we presented detailed analysis of the different families of the defect-free surface states in modulated lattices.

The work was supported by the Australian Research Council through Discovery and Centre of Excellence projects.

\end{sloppy}

\begin{thebibliography}{10}

\bibitem{Davidson:1996:SurfaceStates}
S.~G. Davidson and M. Steslicka, {\em {Basic theory of surface states}} (Oxford
  Science Publications, New York, 1996).

\bibitem{Malkova:2007-45305:PRB}
N. Malkova and C.~Z. Ning, Phys. Rev. B {\bf 76}, 045305 (2007).

\bibitem{Tamm:1932-849:ZPhys}
I.~E. Tamm, Z. Phys. {\bf 76}, 849 (1932).

\bibitem{Shockley:1939-317:PREV}
W. Shockley, Phys. Rev. {\bf 56}, 317 (1939).

\bibitem{Yeh:1977-423:JOS}
P. Yeh, A. Yariv, and C.~S. Hong, J. Opt. Soc. Am. {\bf 67}, 423 (1977).

\bibitem{Yeh:1978-104:APL}
P. Yeh, A. Yariv, and A.~Y. Cho, Appl. Phys. Lett. {\bf 32}, 104 (1978).

\bibitem{Christodoulides:2003-817:NAT}
D.~N. Christodoulides, F. Lederer, and Y. Silberberg, Nature {\bf 424}, 817
  (2003).

\bibitem{Suntsov:2006-63901:PRL}
S. Suntsov {\it et~al.}, Phys. Rev. Lett. {\bf 96}, 063901 (2006).

\bibitem{Makris:2005-2466:OL}
K.~G. Makris {\it et~al.}, Opt. Lett. {\bf 30}, 2466 (2005).

\bibitem{Kartashov:2006-73901:PRL}
Y.~V. Kartashov, V.~A. Vysloukh, and L. Torner, Phys. Rev. Lett. {\bf 96},
  073901 (2006).

\bibitem{Siviloglou:2006-5508:OE}
G.~A. Siviloglou {\it et~al.}, Opt. Express {\bf 14}, 5508 (2006).

\bibitem{Smirnov:2006-2338:OL}
E. Smirnov {\it et~al.}, Opt. Lett. {\bf 31}, 2338 (2006).

\bibitem{Rosberg:2006-83901:PRL}
C.~R. Rosberg {\it et~al.}, Phys. Rev. Lett. {\bf 97}, 083901 (2006).

\bibitem{Molina:2005-35404:PRB}
M.~I. Molina, Phys. Rev. B {\bf 71}, 035404 (2005).

\bibitem{Longhi:2006-243901:PRL}
S. Longhi {\it et~al.}, Phys. Rev. Lett. {\bf 96}, 243901 (2006).

\bibitem{Eisenberg:2000-1863:PRL}
H.~S. Eisenberg, Y. Silberberg, R. Morandotti, and J.~S. Aitchison, Phys. Rev.
  Lett. {\bf 85}, 1863 (2000).

\bibitem{Garanovich:2006-66609:PRE}
I.~L. Garanovich, A.~A. Sukhorukov, and Yu.~S. Kivshar, Phys. Rev. E {\bf 74},
  066609 (2006).

\bibitem{Iyer:2007-3212:OE}
R. Iyer {\it et~al.}, Opt. Express {\bf 15}, 3212 (2007).

\bibitem{Longhi:2005-2137:OL}
S. Longhi, Opt. Lett. {\bf 30}, 2137 (2005).

\bibitem{Somekh:1973-46:APL}
S. Somekh {\it et~al.}, Appl. Phys. Lett. {\bf 22}, 46 (1973).

\bibitem{Kartashov:2005-1378:OL}
Y.~V. Kartashov, L. Torner, and D.~N. Christodoulides, Opt. Lett. {\bf 30},
  1378 (2005).

\bibitem{Garanovich:2005-5704:OE}
I.~L. Garanovich, A.~A. Sukhorukov, and Yu.~S. Kivshar, Opt. Express {\bf 13},
  5704 (2005).

\bibitem{Rosberg:2006-1498:OL}
C.~R. Rosberg {\it et~al.}, Opt. Lett. {\bf 31}, 1498 (2006).

\bibitem{Kivshar:1994-2536:PRE}
Yu.~S. Kivshar and S.~K. Turitsyn, Phys. Rev. E {\bf 49}, R2536 (1994).

\bibitem{Dunlap:1986-3625:PRB}
D.~H. Dunlap and V.~M. Kenkre, Phys. Rev. B {\bf 34}, 3625 (1986).

\end{thebibliography}
\end{document}